Specific features of the magnetic-field dependences of electrical resistivity in Bi–Mn solid solutions with low Mn content

A.V. Terekhov[1], V.M. Yarovyi[1], Yu.A. Kolesnichenko[1], K. Rogacki[2], E. Lähderanta[3], E.V. Khristenko[1], and A.L. Solovjov[1,2,3]

[1] *Verkin Institute for Low Temperature Physics and Engineering, Kharkov, Ukraine, 61103*

[2] *W. Trzebiatowski Institute for Low Temperatures & Structure Research PAS, P.O. Box 1410, 50–950, Wroclaw, Poland*

[3] *Lappeenranta University of Technology, School of Engineering Science, 53850 Lappeenranta, Finland*

e-mail: terekhov@ilt.kharkov.ua, terekhov.andrii@gmail.com

**Introduction**

Bismuth has long attracted the attention of researchers due to its wide range of unusual physical properties, which have been the subject of intensive study for more than a century. A number of fundamental physical effects were first observed in bismuth, including Shubnikov–de Haas and de Haas–van Alphen oscillations, large positive magnetoresistance, cyclotron resonance in metals, oscillating magnetostriction, microwave absorption phenomena and several size-dependent effects [1, 2]. For many years, it was thought that bulk bismuth was incapable of superconductivity. However, this view was revised when a group of Indian scientists reported the discovery of a superconducting transition in single-crystal bismuth at ultra-low temperatures ($T_c \approx 0.5$ mK) [3].

The numerous compounds and alloys that contain bismuth are equally intriguing because they exhibit a variety of remarkable properties. For instance $Bi_2Sr_2Ca_2Cu_3O_{10+x}$ (BiSCCO, Bi-2223) is a high-temperature superconductor with a critical temperature of around 110 K [4]; the $Bi_{0.88}Sb_{0.12}$ alloy exhibits an anomalous magnetoresistance effect [5]; $Bi_{0.9}Sb_{0.1}$ is a strongly spin-orbit-coupled insulator characterised by an odd number of Dirac points and a non-trivial topological phase [6]; $Na_3Bi$ hosts three-dimensional Dirac fermions with linear dispersion in all momentum directions [7]; and Bi–Mn solid solutions occupy a unique position among ferromagnetic materials and permanent magnets, with a Curie temperature of around 640 K and a high coercive force at room temperature [8-9]. However, while the magnetic properties of Bi–Mn solid solutions have been extensively investigated, their electronic transport behaviour has received much less attention.

It should be noted that the unusual transport and magnetic properties of bismuth-based compounds can be considered in the context of the interaction between different subsystems of elementary excitations, which causes anomalous behavior of physical characteristics also in other modern materials [10–14]. In general, the study of magnetotransport in bismuth-based compounds not only provides insight into the fundamental physics of strongly correlated systems, but also joins these materials to a broader class of low-dimensional and topologically non-trivial structures. At the same time, a new branch of electronics has recently emerged called valleytronics. This is supposed to control degrees of freedom such as valleys and local maxima/minima in the valence or conduction band of multi-valley semiconductors and semimetals [15, 16, 17]. There have even been proposals to create qubits for quantum computers based on valleys [18]. The list of materials for valleytronics includes bismuth [19, 20]. For this reason, a comprehensive study of bismuth and materials based on it is promising.

Several years ago, our group conducted a detailed study of the electrical resistance behaviour of solid solutions of $Bi_{95.69}Mn_{3.69}Fe_{0.62}$ and $Bi_{88.08}Mn_{11.92}$, with the aim of addressing the knowledge gap concerning the transport properties of Bi–Mn solid solutions [21-24]. We discovered giant positive magnetoresistance, as well as anomalies in the temperature dependence of electrical resistance in a magnetic field, in these materials. This behaviour was explained within the framework of a two-band conductivity model combined with the influence of internal magnetism. However, the magnetic-field dependence of the electrical resistance of $Bi_{88.08}Mn_{11.92}$ at different temperatures had not yet been studied, despite the fact that such data could provide important insights into the role of manganese concentration in anomalous magnetotransport. Addressing this issue was the main objective of the present study.

## Samples and experimental method

High-purity bismuth and manganese (≥99.999%) were used as the starting materials. The synthesis and subsequent crystal growth were performed in graphitized quartz ampoules with a diameter of 16–18 mm, evacuated to a residual pressure of ~$10^{-2}$ Pa. The samples were grown by the Bridgman method at 630 K with a translation rate of 1.5 mm/h. These growth conditions allow the formation of both single crystals and textured polycrystals. Temperature control within ±0.5 °C was achieved using a RIF-101 controller. The resulting ingots were cylindrical, and specimens for electrical-resistance measurements were prepared by cutting parallelepiped-shaped pieces (approximately 7 × 2 × 2 mm) along the cylinder base, i.e., nominally perpendicular to the c-axis.

A comprehensive structural analysis of this material was previously reported in Ref. [22]. It was established that $Bi_{95.69}Mn_{3.69}Fe_{0.62}$ is a textured polycrystalline material in which a bismuth matrix contains inclusions of the magnetic α-BiMn phase. We assume that the same situation applies to $Bi_{88.08}Mn_{11.92}$, although this material contains a much larger number of α-BiMn magnetic-phase inclusions.

Magnetoresistance was measured using a standard four-probe configuration on an automated Quantum Design PPMS system. Electrical contacts for the current and voltage leads were prepared using silver paste. Measurements were performed in the alternating-current mode ($I$ = 15 mA, $f$ = 19 Hz), with the current applied along the longest dimension of the sample. Data were collected up to 90 kOe at 5, 10, 20, 30, 40, 60, 80, 100, 120, 150, 200 and 300 K for both orientations, $H \perp I$ and $H \parallel I$. A superconducting solenoid was used to generate the permanent magnetic field during the measurements.

## Experimental results

In this section, we examine the magnetic-field dependence of the magnetoresistance of $Bi_{88.08}Mn_{11.92}$ at different temperatures for the two field orientations, $H \perp I$ and $H \parallel I$. The resulting magnetoresistance behavior is then compared with that previously reported for the $Bi_{95.69}Mn_{3.69}Fe_{0.62}$, which contains less manganese and therefore a reduced amount of the magnetic α-BiMn phase.

## Magnetoresistance for the $H \perp I$ configuration

Figure 1 shows the field dependences of the magnetoresistance of $Bi_{88.08}Mn_{11.92}$, plotted in relative units as $MR = \Delta\rho/\rho_0 * 100\% = (\rho_H - \rho_0)/\rho_0 * 100\%$, measured at fixed temperatures of 5, 10, 20, 30, 40, 60, 80, 100, 120, 150, 200, and 300 K for the $H \perp I$ configuration in magnetic fields up to 90 kOe. As the magnetic field increases, the magnetoresistance increases monotonically in magnitude and remains positive across the entire field range. The behavior of the field dependences differs qualitatively across different temperature intervals. Thus, within the 5–80 K range (see Fig. 1a), the curves measured at different temperatures overlap as the magnetic field increases up to 22 kOe, and the MR values increase monotonically. In stronger magnetic fields, the field dependences diverge and the magnetoresistance magnitude increases with temperature. Completely opposite behavior of the magnetoresistance is observed within the temperature range of 100–300 K (see Fig. 1b). Firstly, within this range, the magnetoresistance decreases as the temperature increases, and the magnetoresistance curves no longer coincide. The maximum magnetoresistance was observed at 100 K, with an $MR$ value of approximately 3170 % being achieved in a magnetic field of 90 kOe. At temperatures below 100 K, particularly within the helium temperature range, magnetoresistance in fields exceeding 90 kOe tends to approach saturation. Above 150 K, however, the magnetoresistance in strong magnetic fields increases linearly and there is no tendency for the curves to saturate.

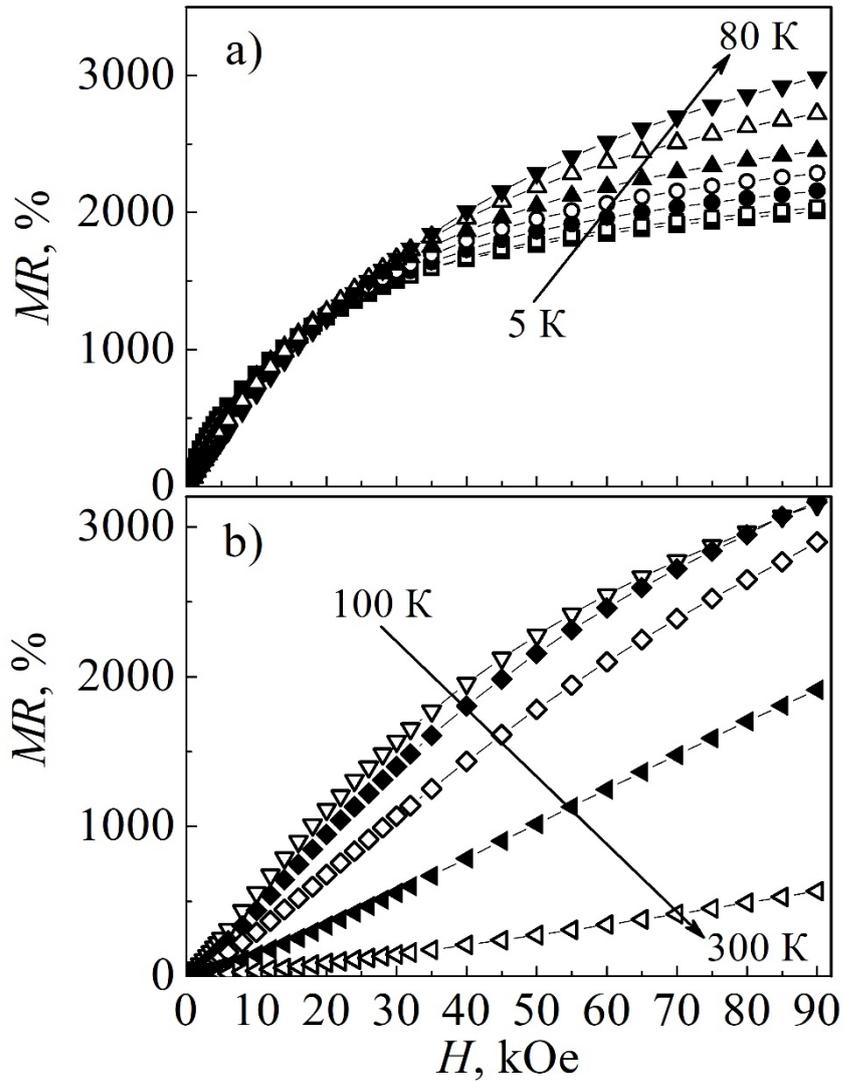

Fig. 1. Field dependences of the magnetoresistance of $Bi_{88.08}Mn_{11.92}$ expressed in relative units $MR = \Delta\rho/\rho_0*100\% = (\rho_H-\rho_0)/\rho_0*100\%$ for the $H \perp I$ configuration, measured in the 5–80 K range (Fig. 1a) at fixed temperatures of 5, 10, 20, 30, 40, 60, and 80 K, as well as in the 100–300 K range (Fig. 1b) at fixed temperatures of 100, 120, 150, 200, and 300 K.

Figure 2 shows the magnetic field dependences of the magnetoresistance of $Bi_{95.69}Mn_{3.69}Fe_{0.62}$ (Fig. 2a, c, e) from our earlier work [23] and $Bi_{88.08}Mn_{11.92}$ (Fig. 2b, d, f) for comparison. The data are shown in relative units, $\Delta\rho/\rho_0 = (\rho_H-\rho_0)/\rho_0$, for the $H \perp I$ configuration and were measured at temperatures of 5 K (Fig. 2a, b), 80 K (Fig. 2c, d) and 150 K (Fig. 2e, f). For comparison, the inset to Fig. 2a also shows literature data for pure bismuth [25] measured over the same magnetic field range as that used for $Bi_{88.08}Mn_{11.92}$ and $Bi_{95.69}Mn_{3.69}Fe_{0.62}$. It should be noted that the investigated Bi–Mn solid solutions consist of a bismuth matrix with inclusions of the magnetic α-BiMn phase. The proportion of this phase is higher in materials with a greater manganese concentration.

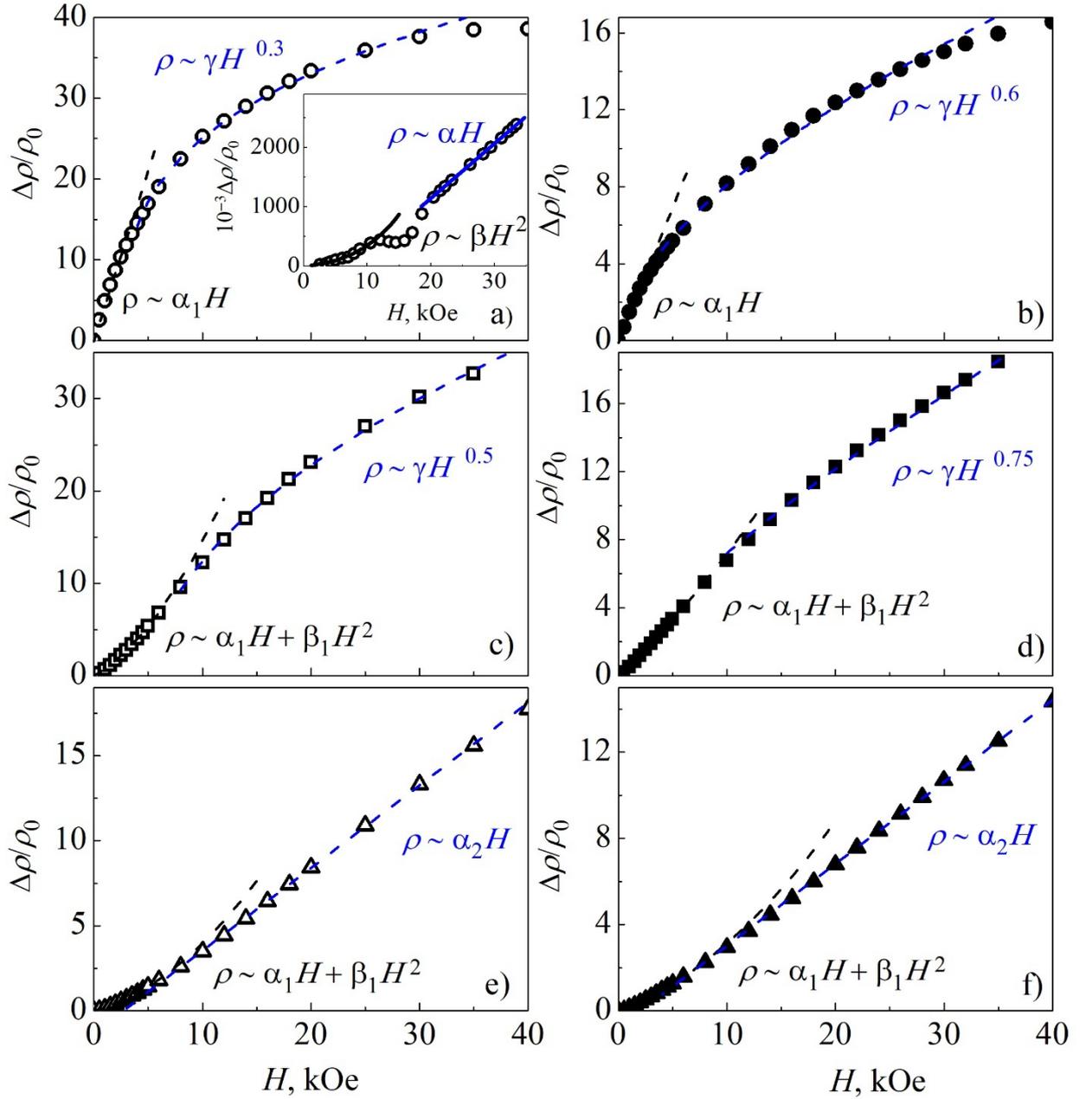

Fig. 2. Field dependences of the magnetoresistance of $Bi_{95.69}Mn_{3.69}Fe_{0.62}$ (Fig. 2 a, c, e) [23] and $Bi_{88.08}Mn_{11.92}$ (Fig. 2 b, d, f) expressed in relative units $\Delta\rho/\rho_0 = (\rho_H-\rho_0)/\rho_0$ for the $\boldsymbol{H} \perp \boldsymbol{I}$ configuration, measured at temperatures of 5 K (Fig. 2a, b), 80 K (Fig. 2c, d), and 150 K (Fig. 2e, f). For comparison, the inset of Fig. 2a shows the data from the literature for pure bismuth [25], which was measured over the same magnetic field range as $Bi_{88.08}Mn_{11.92}$ and $Bi_{95.69}Mn_{3.69}Fe_{0.62}$.

As can be seen, for both materials, the dependences of $\Delta\rho/\rho_0(H)$ at 5 K is linear in weak magnetic fields, whereas in stronger fields, it exhibits weak power-law behavior (Fig. 2a and 2b). On the other hand, the magnetoresistance in pure bismuth displays a quadratic response in weak magnetic fields and a linear one in strong ones [25–27]. At the same time, as the concentration of manganese increases and we move from $Bi_{95.69}Mn_{3.69}Fe_{0.62}$ to $Bi_{88.08}Mn_{11.92}$, the exponent roughly doubles. The dependence of $\Delta\rho/\rho_0$ on H for both compositions at 80 K in weak magnetic fields is given by $\Delta\rho/\rho_0 \sim \alpha_1 H + \beta_1 H^2$, while in stronger fields the dependence evolves into a weak power-law behavior (see Fig. 2c, d). The exponent is higher for the $Bi_{88.08}Mn_{11.92}$ sample, which has a higher manganese concentration. As the temperature increases up to 150 K, the dependence of $\Delta\rho/\rho_0$ on $H$ remains the same in weak magnetic fields, whereas in strong fields it becomes linear (Fig. 2c and 2d). The same dependences are also observed for both materials at 300 K, but over a wider range of magnetic fields (Fig. 3a and 3b). Thus, it can be concluded that the dependencies are similar for $Bi_{95.69}Mn_{3.69}Fe_{0.62}$ and $Bi_{88.08}Mn_{11.92}$ at temperatures above 150 K. However, at lower temperatures (5 K and 80 K), they differ in strong magnetic fields (above 25 kOe) in terms of the exponent, which is closer to 1 for $Bi_{88.08}Mn_{11.92}$ with the higher manganese concentration.

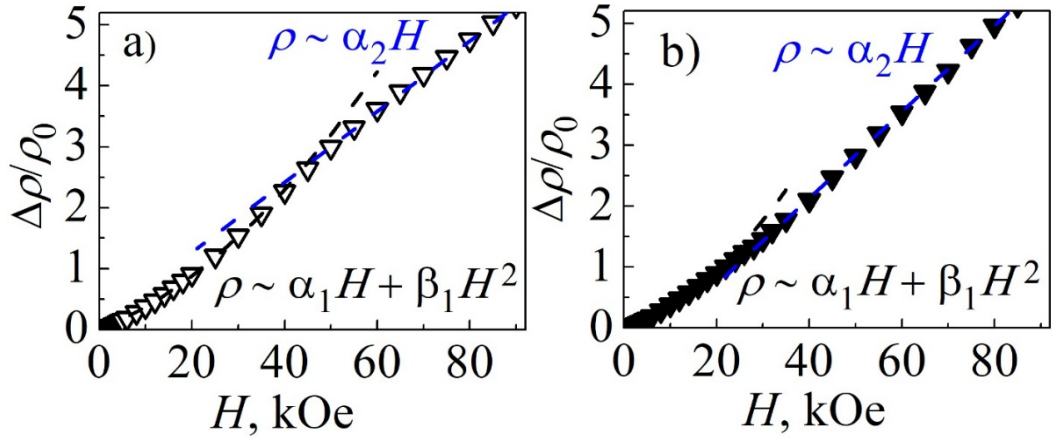

Fig. 3. The magnetic field dependences of the magnetoresistance of $Bi_{95.69}Mn_{3.69}Fe_{0.62}$ (Fig. 3 a) [23] and $Bi_{88.08}Mn_{11.92}$ (рис. 3 b) expressed in relative units $\Delta\rho/\rho_0 = (\rho_H - \rho_0)/\rho_0$ for the $\boldsymbol{H} \perp \boldsymbol{I}$ configuration, measured at temperature of 300 K.

At the same time, for the investigated Bi–Mn solid solutions, the dependence $\Delta\rho/\rho_0(H)$ at 5 K in strong magnetic fields follows weak power-law behavior, whereas in pure bismuth it is linear. In Ref. [28], the material $Bi_{50.8}Mn_{49.2}$ with an even higher manganese concentration was investigated. This material consisted of a ferromagnetic α-BiMn matrix with virtually no bismuth inclusions. In this case, at 5 K, in contrast to the $Bi_{95.69}Mn_{3.69}Fe_{0.62}$ and $Bi_{88.08}Mn_{11.92}$ solid solutions with lower manganese content investigated in our work, a linear magnetoresistance was observed in magnetic fields above 10 kOe, with no tendency toward saturation up to 90 kOe. Meanwhile, negative magnetoresistance was frequently observed at temperatures above 100 K, reaching saturation in magnetic fields of up to 30 kOe. At the same time, the magnetoresistance remained positive over the entire temperature range in the materials investigated in our work, as well as in pure bismuth. Thus, it can be argued that, in both $Bi_{88.08}Mn_{11.92}$ and $Bi_{95.69}Mn_{3.69}Fe_{0.62}$, the transport current predominantly flows through the bismuth matrix and that the conductivity behavior in a magnetic field is primarily governed by the bismuth. The differences between the magnetoresistance dependencies of our materials and pure bismuth are attributed to the influence of magnetism associated with α-BiMn phase inclusions on charge transport. An increase in the concentration of this phase results in an increase in the exponent value in magnetic fields above 25 kOe at temperatures below 150 K. At the same time, for T > 150 K, the dependences of $\Delta\rho/\rho_0(H)$ for these two solid solutions are identical.

### Magnetoresistance for the *H // I* configuration

Figure 4 shows the dependence of the magnetoresistance on the magnetic field, plotted in relative $MR = \Delta\rho/\rho_0 * 100\% = (\rho_H - \rho_0)/\rho_0 * 100\%$, measured at fixed temperatures of 5, 10, 20, 30, 40, 60, 80, 100, 120, 150, 200 and 300 K for the *H // I* configuration in magnetic fields up to 90 kOe.

The behavior of the dependences differs qualitatively in different temperature ranges. In the 5–80 K range (Fig. 4a), with increasing magnetic field up to approximately 26 kOe, the curves corresponding to different temperatures nearly coincide, and *MR* increases monotonically. In stronger magnetic fields, the dependencies of *MR(H)* begin to diverge. Subsequently, the magnetoresistance reaches a maximum, after which it decreases as the field increases further. As shown in the inset to Fig. 4a, the maximum values on the *MR(H)* curves increase with temperature in an approximately linear manner. As the temperature increases, the high-field portion of the *MR(H)* curves shifts toward lower values. Meanwhile, the magnetoresistance remains positive across the entire magnetic field range investigated. The opposite behavior is observed within the 100–300 K range (Fig. 4b). Firstly, within this temperature range, the magnetoresistance decreases as the temperature increases, and the *MR(H)* curves no longer coincide within weak magnetic fields. Secondly, unlike the behavior at *T* < 100 K, the magnetoresistance increases monotonically with a tendency towards saturation in strong magnetic fields. The maximum value was recorded at 100 K: *MR* ≈ 380% in a magnetic field of 90 kOe.

Figure 5 shows the magnetic field dependence of the magnetoresistance of $Bi_{95.69}Mn_{3.69}Fe_{0.62}$ (Fig. 5a, c, e, g) from our previous study [23], as well as the results obtained in this study for $Bi_{88.08}Mn_{11.92}$ (Fig. 5b, d, f, h). These are plotted in relative units, $\Delta\rho/\rho_0 = (\rho_H - \rho_0)/\rho_0$, for the *H // I* configuration. Measurements were taken at temperatures of 5 K (Fig. 5a, b), 80 K (Fig. 5c, d), 150 K (Fig. 5e, f) and 300 K (Fig. 5g, h).

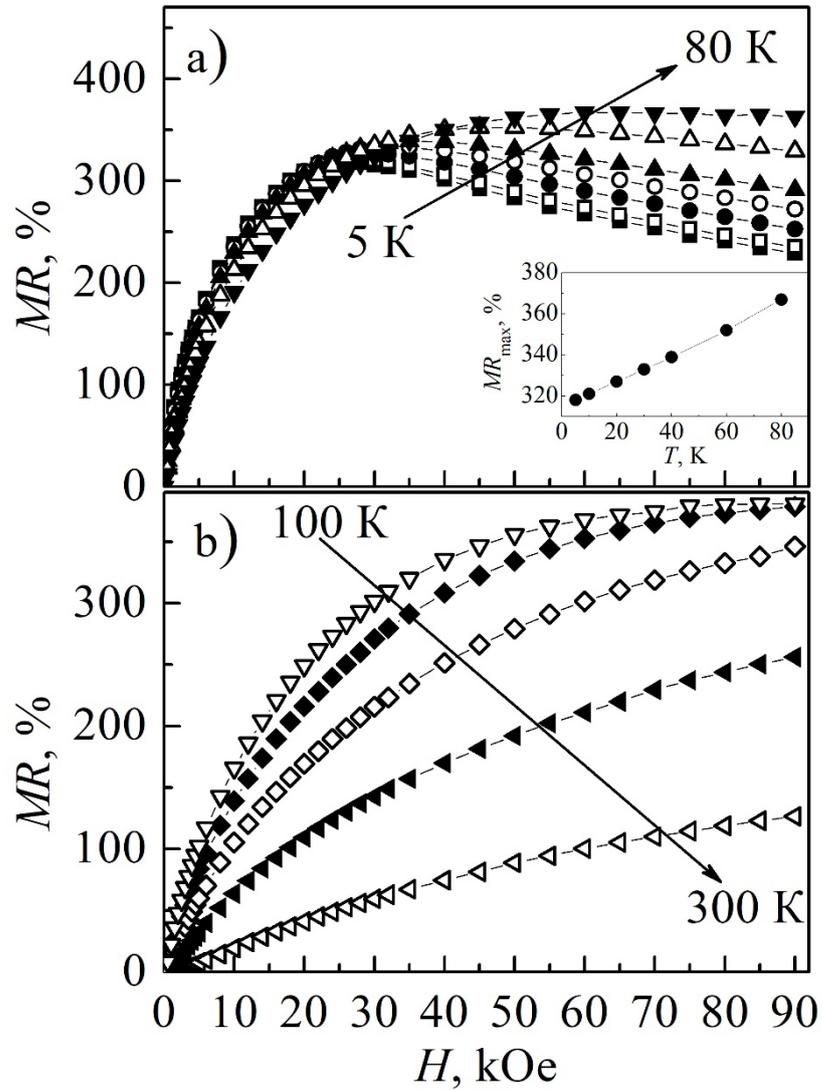

Fig. 4. The magnetic field dependences of the magnetoresistance of $Bi_{88.08}Mn_{11.92}$ expressed in relative units $MR = \Delta\rho/\rho_0*100\% = (\rho_H-\rho_0)/\rho_0*100\%$ for the ***H** // **I*** configuration, measured in the 5–80 K range (Fig. 4a) at fixed temperatures of 5, 10, 20, 30, 40, 60, and 80 K, as well as in the 100–300 K range (Fig. 4b) at fixed temperatures of 100, 120, 150, 200, and 300 K..

Both materials exhibit weak power-law behavior for the dependence $\Delta\rho/\rho_0(H)$ at 5 K in weak magnetic fields, with a higher exponent for $Bi_{88.08}Mn_{11.92}$. The dependence is linear for $Bi_{95.69}Mn_{3.69}Fe_{0.62}$ in stronger magnetic fields, whereas for $Bi_{88.08}Mn_{11.92}$ it follows a power-law behavior with a subsequent tendency toward saturation. At the same time, the exponent decreases as the magnetic field increases. The $\Delta\rho/\rho_0(H)$ dependences at 80 K are similar for both compositions. The dependence of $\Delta\rho/\rho_0(H)$ in weak magnetic fields follows a power law, with a higher exponent in materials with greater manganese content. The magnetoresistance in stronger magnetic fields becomes linearly dependent on the field. The $\Delta\rho/\rho_0(H)$ dependence for $Bi_{95.69}Mn_{3.69}Fe_{0.62}$ at 150 K, follows a power-law behavior up to 30 kOe. For $Bi_{88.08}Mn_{11.92}$, the field dependence of the magnetoresistance is linear up to 5 kOe, after which it exhibits power-law behavior in the 5–30 kOe range. The exponent is slightly higher than that observed in materials with lower manganese content. The dependences of $\Delta\rho/\rho_0$ on $H$ at 300 K are similar for both materials: $\Delta\rho/\rho_0 \sim \alpha_1 H + \beta_1 H^2$ in weak magnetic fields, whereas at higher fields the dependence becomes linear. Let us compare our results with the literature data [28] for the high-manganese-content compound $Bi_{50.8}Mn_{49.2}$, which consists of a ferromagnetic α-BiMn matrix through which the transport current flows. This is in contrast to our case, where the matrix is assumed to be bismuth with inclusions of the magnetic α-BiMn phase, and where no percolative conduction paths between these inclusions are present. A monotonically increasing, positive $\Delta\rho/\rho_0(H)$ dependence, without a maximum and without saturation in magnetic fields up to 90 kOe was observed in the $Bi_{50.8}Mn_{49.2}$ sample for the $H // I$ configuration at temperatures of 5–70 K. This behavior differs from the field dependences observed in the $Bi_{95.69}Mn_{3.69}Fe_{0.62}$ and $Bi_{88.08}Mn_{11.92}$ solid solutions with lower manganese content in the same temperature range. Thus, it can be argued that in both $Bi_{88.08}Mn_{11.92}$ and $Bi_{95.69}Mn_{3.69}Fe_{0.62}$ the transport current flows through the bismuth matrix, and the conductivity behavior in a magnetic field is governed by bismuth. The differences between the magnetoresistance dependences of our materials and pure bismuth can be attributed to the influence of the magnetism of the α-BiMn phase inclusions on the charge transport.

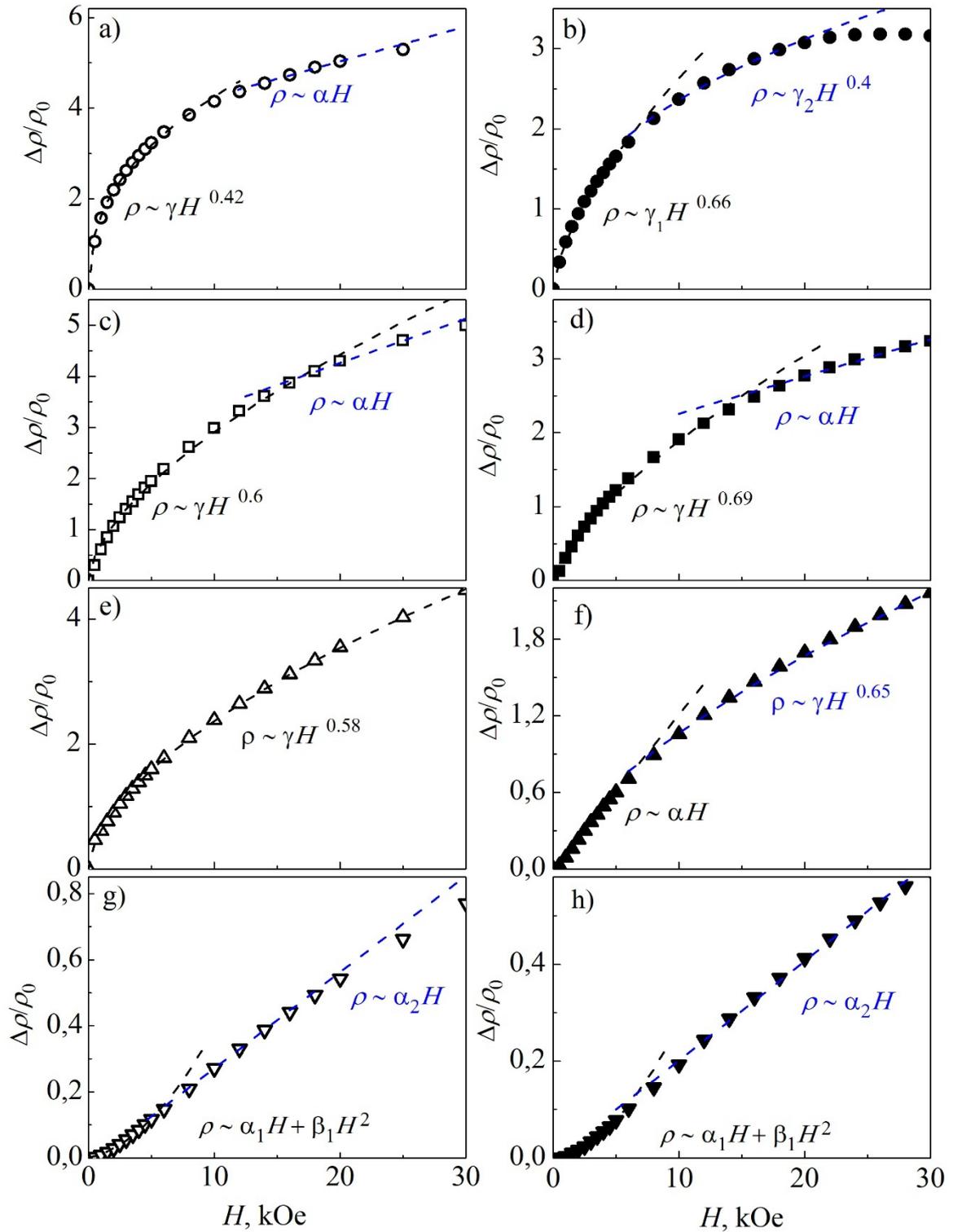

Fig. 5. The magnetic field dependences of the magnetoresistance of $Bi_{95.69}Mn_{3.69}Fe_{0.62}$ (Fig. 5 a, c, e, g) [23] and $Bi_{88.08}Mn_{11.92}$ (Fig. 5 b, d, f, h) expressed in relative units $\Delta\rho/\rho_0 = (\rho_H - \rho_0)/\rho_0$ for the $\boldsymbol{H} // \boldsymbol{I}$ configuration, measured at temperature of 5 K (Fig. 5 a, b), 80 K (Fig. 5 c, d), 150 K (Fig. 5 e, f) and 300 K (Fig. 5 g, h).

## Discussion of results

For comparison, Table 1 presents the data on the maximum magnetoresistance at 90 kOe for $Bi_{95.69}Mn_{3.69}Fe_{0.62}$ and $Bi_{88.08}Mn_{11.92}$ for magnetic field orientations of $H \perp I$ and $H // I$. It is clearly seen that with increasing fraction of the magnetic α-BiMn phase, the magnetoresistance decreases markedly. It should be noted that, as the manganese content increases, the magnetoresistance decreases by almost half for the $H // I$ configuration and by around 20% for the $H \perp I$ configuration. The possible origin of this behavior will be discussed below.

Table 1. Maximum magnetoresistance values $MR = \Delta\rho/\rho_0*100\% = (\rho_H - \rho_0)/\rho_0*100\%$ for $Bi_{95.69}Mn_{3.69}Fe_{0.62}$ and $Bi_{88.08}Mn_{11.92}$ for the magnetic field orientations $H \perp I$ and $H // I$.

| Samples | $MR_{max}$, $H \perp I$, at 90 кOe | $MR_{max}$, $H // I$, at 90 кOe |
|---|---|---|
| $Bi_{95.69}Mn_{3.69}Fe_{0.62}$ | 3877 % | 742 % |
| $Bi_{88.08}Mn_{11.92}$ | 3170 % | 380 % |

As discussed above, the behavior of magnetic field dependences for these materials also differs significantly, especially in the temperature range below 100 K. As the temperature approaches room temperature, the dependences become similar. What happens at temperatures below 100 K? According to the literature on the magnetic properties of Bi–Mn solid solutions [9, 29] and our study of $Bi_{95.69}Mn_{3.69}Fe_{0.62}$ [22], a change in magnetic structure, namely a reorientation of the Mn magnetic moments, may occur in such materials below 100 K. The latter change their orientation from perpendicular to parallel with respect to the *c* crystallographic axis. Below, we consider how this may affect the magnetoresistance of the bismuth-manganese solid solutions examined in this study.

According to literature data [30], the Fermi energy in bismuth is extremely small, amounting to only a few hundredths of an electron volt. Therefore, even relatively weak external perturbations, such as changes in temperature, the magnetic field or magnetic interactions, or mechanical strain, can modify the mutual overlap of the electron and hole pockets in *k*-space. Consequently, the transport properties, particularly the electrical conductivity, are also altered. Under certain conditions, when the electron and hole Fermi surfaces almost overlap, a transition from a metallic to a semiconducting state may occur.

Bismuth is a nearly compensated semimetal, meaning that the concentrations of electrons and holes within it are almost equal. As noted in Ref. [30], its Fermi surface comprises one hole pocket situated at the *T*-point and three electron pockets positioned at the *L*-points within the first Brillouin zone of the rhombohedral lattice (Fig. 6). In the absence of a magnetic field, these electron pockets are equivalent and correspond to three degenerate valleys (*e*1, *e*2, *e*3) (Fig. 6).

At the same time, the pronounced anisotropy of the electron pocket shapes (see Fig. 6) results in valley degeneracy being lifted under an applied magnetic field [19, 20]. Small energy gaps ($E_g \approx 15{,}3$ meV) between the valence and conduction bands are observed in the vicinity of the *L*-points (Fig. 7). When the magnetic field is oriented along either the binary or bisectrix axis, the energy gap initially decreases. Upon reaching critical values, a level crossing occurs [31, 32]. Further increases in the magnetic field lead to the electron and hole subbands diverging in opposite directions. In the limiting case, this can cause the electronic state of the material to change from semimetallic to semiconducting.

The features of the electrical resistance dependences identified in the present work for the investigated solid solutions, as compared with pure bismuth, are highly likely to be associated with an additional contribution from magnetic effects arising from the presence of the α-BiMn phase. The magnetic order of this phase may alter the extent to which the electron and hole regions of the Fermi surface overlap, as reflected in the material's transport properties. Additionally, the substantially greater change in magnetoresistance magnitude with decreasing manganese concentration for the ***H // I*** configuration, compared to the ***H ⊥ I*** configuration, may be attributed to the significantly higher magnetisation of the α-BiMn phase for the ***H // I*** configuration (due to the reorientation of the Mn magnetic moments below 100 K). A change in the angle between the magnetic field direction and the crystallographic axes can additionally affect both the overlap of the electron and hole regions and the magnetic configuration of α-BiMn, leading to deviations of the resistance behavior from the $H = 0$ case. As mentioned previously, the most significant changes to the magnetic structure (i.e. the spin-reorientation transition) of the Bi–Mn solid solutions occur below 100 K. This is where the greatest differences in the magnetoresistance behavior of $Bi_{95.69}Mn_{3.69}Fe_{0.62}$ and $Bi_{88.08}Mn_{11.92}$ are observed.

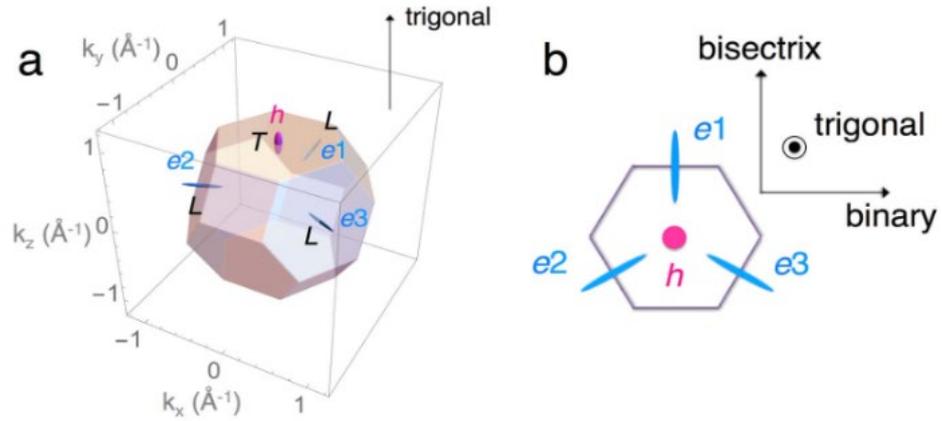

Fig. 6 (a) Three-dimensional schematic pictures of the first Brillouin zone and the Fermi surfaces of bismuth. Three valleys of electron pockets, labelled $e1$, $e2$ and $e3$ (blue), are located at $L$-points, and one valley of hole pockets, labelled h (purple), is located at the $T$-point. (b) Diagram of Fermi surfaces projected onto a binary bisector plane [30].

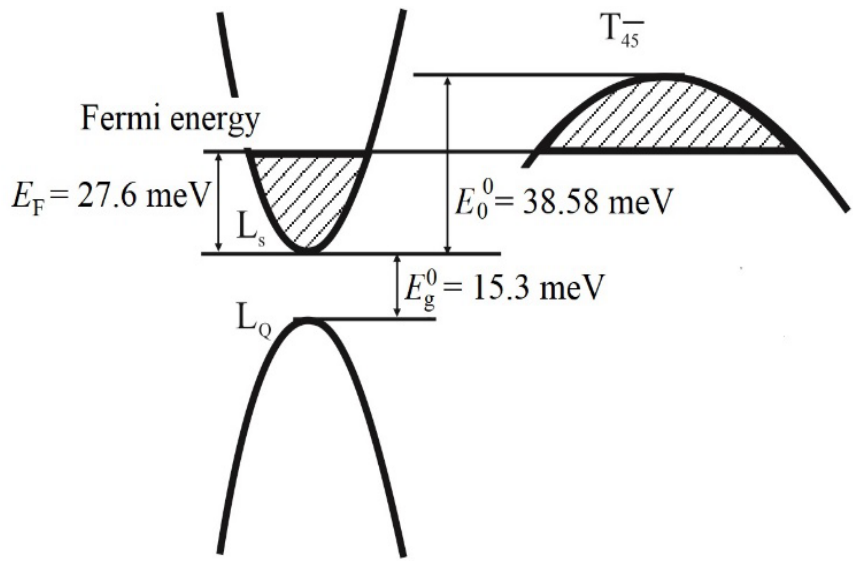

Fig. 7. The energy position of the extrema of the valence band and conduction band at points $L$ and $T$ of the Brillouin zone of a bismuth crystal in the vicinity of the Fermi energy at $T = 4.2$ K.

In our earlier work, we observed a maximum in the field dependence of the magnetoresistance of $Bi_{95.69}Mn_{3.69}Fe_{0.62}$ at helium temperature for magnetic field orientations of $\mathbf{H} \perp \mathbf{I}$ and $\mathbf{H} \parallel \mathbf{I}$. At that time, it was suggested that the quantum limit might be reached in this material by analogy with pure bismuth [33–35], i.e. when only one Landau level remains below the Fermi level in the conduction band. In a semimetal, this results in charge carriers «flowing» between the hole and electron bands. As a result, the charge carrier concentration increases, and the electrical resistivity decreases [34]. At the same time, detailed studies of the magnetoresistance of $Bi_{88.08}Mn_{11.92}$ carried out in this study have shown that maxima in the field dependence of *MR* are observed up to 80 K. This behavior cannot be easily explained by quantum effects and is more likely to be associated with changes in the magnetic order of manganese in the α-BiMn phase.

In a subsequent study, detailed investigations of the magnetic properties of $Bi_{88.08}Mn_{11.92}$ are planned in order to verify this assumption.

### Conclusions

1. The field dependence of the relative magnetoresistance of textured polycrystalline $Bi_{88.08}Mn_{11.92}$ was investigated for the first time in two configurations: with the magnetic field perpendicular to the transport current, and with it parallel. The results were then compared with those for $Bi_{95.69}Mn_{3.69}Fe_{0.62}$, which has a lower manganese concentration.

2. It is shown that the field dependences of the magnetoresistance for $Bi_{88.08}Mn_{11.92}$ differ significantly from those for $Bi_{95.69}Mn_{3.69}Fe_{0.62}$ below 100 K, becoming approximately similar as the temperature approaches room temperature.

3. The maximum values of the relative magnetoresistance of $Bi_{88.08}Mn_{11.92}$ in both the $\mathbf{H} \perp \mathbf{I}$ and $\mathbf{H} \parallel \mathbf{I}$ configurations are observed at a magnetic field of 90 kOe and a temperature of 100 K. The respective values are 3170 % and 380 %. These values are significantly lower than those obtained for the solid solution with a lower manganese content ($Bi_{95.69}Mn_{3.69}Fe_{0.62}$): 3877% and 742%.

4. It is suggested that the different behavior of the field dependences for $Bi_{88.08}Mn_{11.92}$ and $Bi_{95.69}Mn_{3.69}Fe_{0.62}$ may be attributed to the different content of the magnetic α-BiMn phase. More generally, the anomalous magnetoresistance behavior compared to pure bismuth may be associated with the influence of the internal magnetism of the α-BiMn inclusions on the conductive properties of charge carriers within the bismuth matrix of the studied solid solutions.

## ACKNOWLEDGMENTS

The authors are grateful to **professor** Yu.G. Naidyuk and **DSc** I.V. Zolochevskii for their useful discussions and remarks, which were helpful to the process of finalizing this article. The work was supported by the National Academy of Sciences of Ukraine within the F19-5 project. Work was also funded by Office of Naval Research (ONRG) and US National Academy of Sciences (NAS) IMPRESS-U grant via STCU project #7120.